\documentclass[journal]{rmaa}
\def\etal{et\,\,al.\,}
\def\hmpc{$h^{-1}\,$Mpc}
\def\hkpc{$h^{-1}\,$kpc}
\def\hmsun{$h^{-1}\,$M$_\odot$}

\def\mnras{MNRAS}
\def\apj{ApJ}
\def\apjl{ApJ}
\def\apjs{ApJS}
\def\jcap{JCAP}
\def\aap{A\&Ap}
\def\nat{Nature}
\def\aj{AJ}
\def\jcap{JCAP}
\def\pasp{PASP}
\def\araa{ARAA}

\def\spose#1{\hbox to 0pt{#1\hss}}
\def\lta{\mathrel{\spose{\lower 3pt\hbox{$\mathchar"218$}}
     \raise 2.0pt\hbox{$\mathchar"13C$}}}
\def\gta{\mathrel{\spose{\lower 3pt\hbox{$\mathchar"218$}}
     \raise 2.0pt\hbox{$\mathchar"13E$}}}

\usepackage[dvips]{epsfig}
\usepackage{graphicx}
\title{Intragroup dark matter distribution in small groups of halos in a $\Lambda$CDM cosmology }  
\author{H. Aceves, F.J. Tamayo, L. Altamirano-D\'evora,  F.G. Ram\'on-Fox, R. Ca\~nas \& M. Reyes-Ruiz}

\altaffiltext{1}{Instituto de Astronom\'{\i}a, UNAM. Ensenada, M\'exico.}

\fulladdresses{
\item H. Aceves, F.J. Tamayo, L. Altamirano-D\'evora,
  F.G. Ram\'on-Fox, R. Ca\~nas \& M. Reyes-Ruiz: Instituto de
  Astronom\'{\i}a, Universidad Nacional Aut\'onoma de
  M\'exico. Apdo. Postal 106, Ensenada B.~C. 22800 M\'exico
  (aceves@astro.unam.mx).  }

\listofauthors{H. Aceves, F.J. Tamayo, L. Altamirano-D\'evora,  F.G. Ram\'on-Fox, R. Ca\~nas \& M. Reyes-Ruiz}
\indexauthor{Aceves, H.}
\indexauthor{Tamayo, F.J.}
\indexauthor{Altamirano-D\'evora, L.}
\indexauthor{Ram\'on-Fox, F.G.}
\indexauthor{Ca\~nas, R.}
\indexauthor{Reyes-Ruiz, M.}

\shortauthor{Aceves et al.}
\shorttitle{Dark matter in groups}

\resumen{Se estudia la distribuci\'on de materia oscura intragrupal en peque\~nos grupos de halos oscuros de tama\~no gal\'actico en una cosmolog\'{\i}a $\Lambda$CDM. Estos grupos oscuros son identificados utilizando un criterio f\'{\i}sico, y pueden ser representativos de peque\~nos grupos de galaxias. Cuantificamos la cantidad de materia oscura intragrupal y caracterizamos su distribuci\'on. Encontramos que las asociaciones compactas de halos, y las intermedias y mucho menos compactas, tienen perfiles de masa oscura algo planos con pendiendes logar\'{\i}tmica de $\gamma\!\approx 0\!$  y $\approx -0.2$, respectivamente. Concluimos entonces que la materia oscura intragrupo en estos sistemas no sigue la misma distribuci\'on que la de los halos gal\'acticos.
 En grupos intermedios u holgados de halos la materia intragrupal es $\lta 50$\%, mientras que en los compactos $\lta 20$\% dentro del radio del grupo. }

\abstract{
  We study the distribution of intragroup dark matter in small groups of dark matter halos of galaxy-like size in a $\Lambda$CDM cosmology. These groups are identified using a physical criterion and may be an appropriate representation of small galaxy groups. We quantify the amount of intra-group dark matter and characterize its distribution. We find that compact associations of halos, as well as those intermediate and loose groups, have rather flat intragroup dark matter profiles with logarithmic slopes of $\gamma \approx 0$  and $\approx -0.2$, respectively.
 Hence, the intra-group dark matter of these  halo systems does not follow the same cuspy tendency that halos of galaxies have. 
 In intermediate and loose galaxy-size halo associations the intragroup matter tends to be $\lta 50$\% that of the total mass of the group, and in compact associations is $\lta 20$\% within their group radius.
}


\begin{document}
\maketitle
%%%%%%%%%%%%%%%%%%%%%%%%%%%%%%%%%%%%%%%%%%%%%%%%%%%%%%%%%%%%%%%%%%%%%%%%%
%%%%%%%%%%%%%%%%%%%%%%%%%%%%%%%%%%%%%%%%%%%%%%%%%%%%%%%%%%%%%%%%%%%%%%%%%
%\baselineskip=20pt

\section{Introduction}\label{sec:intro}
%%%%%%%%%%%%%%%%%%%%%%%%%%%%%%%%%%%%%%%%%%%%%%%%%%%%%%%%%%%%%%%%%%%%%%%%%
%%%%%%%%%%%%%%%%%%%%%%%%%%%%%%%%%%%%%%%%%%%%%%%%%%%%%%%%%%%%%%%%%%%%%%%%%

The formation of structures is a general characteristic of the gravitational interaction among particles in the Universe, regardless of the cosmological model one uses to describe it. The details of how the growth of structures proceed in a simulation are dependent, however, on the adopted cosmological model. It is an observational fact that a large percentage of galaxies at lower redshifts live in aggregates from small groups to large clusters of galaxies (e.g. Holmberg 1950, Tully 1987, Nolthenius \& White 1987, Eke et al.~2004).

The importance of understanding groups of galaxies and the evolution of galaxies within such environments was the main motivation for early catalogs such as those of Tully~(1980) and Huchra \&  Geller~(1982), and of more recent observational efforts including those that aim {\bf at} determining the mass distribution within groups and clusters (e.g. Eke et al.~2004,  Brough et~al.~2006, Berlind et~al.~2006, Yang~et~al.~2007, Tago et al.~2010, Calvi et al.~2011, Carollo et al.~2012, Williams et~al. 2012, Dom\'{\i}nguez-Romero~et~al.~2012, Tempel~et~al.~2014). 

In regard to works on small groups of galaxies, in particular those of a compact nature (e.g. Hickson~1997, Tovmassian~et~al.~1999, Allam \& Tucker~2000, de Carvalho~et~al.~2005, Niemi et al.~2007, McConnachie et~al.~2008,  Mamon~2008, D\'{\i}az-Gim\'enez et al.~2012), a recurrent topic is the question of the abundance of physical groups and explaining their  ``existence'' given their small crossing times.  Dynamical studies (e.g. Barnes~1989, Athanassoula, Makino \& Bosma 1997, G{\'o}mez-Flechoso  
\& Dom{\'{\i}}nguez-Tenreiro~2001, Aceves \& Vel\'azquez 2002) have shown that compact groups may have a long existence given particular initial conditions. Cosmological simulations (e.g. Diaferio et al.~1994, Governato~et al.~1996, Casagrande \& Diaferio 2006, Sommer-Larsen~2006) have also addressed the question of compact groups, including part of the baryonic physics of gas.
  Comparisons of observed compact associations (CAs) with mock catalogues have  led to the conclusion that a significant fraction (about 30\%) of observed compact groups in observational catalogs are physical systems (e.g. McConnachie et al.~2008, Mamon~2008, D\'{\i}az-Gim\'enez \& Mamon 2010), with percentages differing depending on the specifics of the selection criteria used for constructing the mock catalogs (e.g. Duarte \& Mamon 2014). The problem of the longevity of small compact groups is still an open question. 

The amount and distribution of luminous and dark matter in structures in the universe, from galaxies to clusters, is an important problem and may well serve to discriminate or impose restrictions to cosmological models. Intracluster light observed in, for example, the Coma Cluster and  other clusters (e.g. Zwicky~1951, Gonzalez et~al.~2000) suggest that it may conform from 10 to 50 percent of the total light of such structures. Intragroup diffuse light in some Hickson's compact groups has also been observed, where different fractions have been quoted; e.g.,  for HCG44 is its found about 5 percent of the total light  (Aguerri et~al.~2006),  and for HCG95 and HCG79 fractions of about 11 and 45 percent, respectively, have been found (Da Rocha \& Mendes de Oliveira~2005).

The dark matter content and distribution is more or less well established in large structures such as clusters by observations and analysis of gravitational lensing (e.g. Bartelmann~2010, Newman et al.~2012) among other methods. Sand et al.~(2004) find that the dark matter distribution in some clusters of galaxies is inconsistent with the NFW profile. Studies at the mass range of small groups, $M_g\!\approx\! 10^{13} {\rm M}_\odot$, are almost non-existent due primary to observational limitations. The amount and distribution of dark matter in small groups is important, for instance, for dynamical studies of interactions of galaxies in such environments. 
Several authors have investigated the mass content of galaxy groups with lensing methods (e.g. Hoekstra et al.~2001, Parker et al.~2005, Limousin et~al.~2009, Thanjavur et al.~2010, McKean et al.~2010) thus avoiding  the complications of other methods that depend on the dynamical state of the system or its gas temperature. In particular Thanjavur et al.~(2010) find for groups with a mass close to  $10^{14} {\rm M}_\odot$ that dark matter is distributed in a cuspy manner, while Hoekstra et al.~2001 found a tendency towards an isothermal profile for other set of groups (CNOC2 groups, Carlberg et al.~2001) with a lower mass of $\approx 10^{13} {\rm M}_\odot$.

The purpose of this work is to quantify and characterize the amount of diffuse or intra-group (IG) dark matter in small groups of galaxy-size dark matter halos; i.e. that dark matter not bound to well defined virialized halos,  estimated to be in the mass range of $M\in [10^{11},5\times 10^{12}] h^{-1} {\rm M}_\odot$. This is with the aim of obtaining an estimation of what might one expect in true physical small galaxy groups, since these halos can host normal galaxies.
 We study such distribution of dark matter using a set of five cosmological simulations within the $\Lambda$CDM cosmology ($\S2.1$). Bound groups of halos are identified using a physically motivated algorithm, that leads to an unambiguous identification of groups in our simulations ($\S2.2$). 
 Membership in a small dark group is determined by considering only halos that can host ``normal'' galaxies ($\S2.2.2$).  In the analysis of the simulations we will differentiate, as it is done in observational studies, from compact associations, and intermediate and loose groups by means of the size of the group radius $R_{\rm g}$. 

 The outline of the paper is as follows. In Section~\ref{sec:methods} we describe our simulations and the methods used in this study, such as the algorithm used to determine which halos belong  to a dark group or not. In Section~\ref{sec:results} we present our results regarding the intragroup (IG) amount and distribution of dark matter, as well as the evolution in time of the IG dark matter profile for  compact associations of galaxies. In Section~\ref{sec:final} we provide some final comments on our work.

%%%%%%%%%%%%%%%%%%%%%%%%%%%%%%%%%%%%%%%%%%%%%%%%%%%%%%%%%%%%%%%%%%%%%

\section{Methodology}
\label{sec:methods}
%%%%%%%%%%%%%%%%%%%%%%%%%%%%%%%%%%%%%%%%%%%%%%%%%%%%%%%%%%%%%%%%%%%%%5

\subsection{Cosmological Simulations}
%%%%%%%%%%%%%%%%%%%%%%%%%%%%

Our groups of halos were obtained from a set of five similar cosmological simulations within the $\Lambda$CDM model, each differing from each other in the random seed used to generate the initial conditions.  
 The cosmological parameters taken are consistent with those of the {\sc Wmap7} results (Larson \etal 2011, Table~3), where we took for the matter density $\Omega_m\!=\!0.27$, dark energy density $\Omega_\Lambda\!=\!0.73$, spectral index $n_s\!=\!0.963$, mass fluctuation $\sigma_8\!=\!0.816$  and the Hubble parameter $h\!=\!0.70$.  
Each simulation box has a comoving length of $L\!=\!100\,$\hmpc\/  with $N_p=512^3$ dark matter particles, with each particle having a mass of  $m_p\approx 6\times 10^8$\hmsun.

Initial conditions were generated using 2nd-order Lagrangian perturbation theory (e.g. Crocce, Pueblas \& Scoccimarro 2006) at a redshift of $z\!=\!50$. This value is sufficiently high to avoid the effects of transient modes that result  from a truncation in perturbation theory at redshifts of $z\approx 5$ (e.g. Tatekawa \& Mizuno~2007). For halos in the range of $M\in 10^{10-13} h^{-1}\,$M$_\odot$, it appears that first and second order perturbation methods at $z\!=\!0$ do not make, however, an important difference in halo properties (Knebe et al.~2009).  The initial linear power spectrum density is calculated using the transfer function from the cosmic microwave background code {\sc camb} (Lewis, Challinor \& Lasenby 2000), normalized so that it gives the current mass fluctuation $\sigma_8$ value above.

The $N$-body cosmological simulations were carried out using the publicly available  parallel Tree-PM code {\sc Gadget}2 (Springel~2005). The simulations were run with code  parameters similar to those identified as ``high quality'' (HQ) in the simulations of Crocce~et~al.~(2006): for example, using a softening length of $\varepsilon=20$\hkpc. Since we are not interested in small halo substructures, but more on scales of dark halos of typical normal galaxies with virial radius $\approx 200\,$kpc, we do not expect important  differences in our group finding methods on such parameters as a  function of the softening. We were able to test and verify the previous by re-doing 3 similar simulations but with  $\varepsilon=20$\hkpc.

\subsection{Haloes and Group Identification}
%%%%%%%%%%%%%%%%%%%%%%%%%%%%

\subsubsection{Haloes}
%%%%%%%%%%%%%

 Several halo finders exist and many of them have been recently compared (Knebe \etal 2011), although others newer were naturally excluded at the time of such comparison study (e.g. Elahi,  Thacker \& Widrow 2011, Han \etal 2011). We chose the Amiga Halo Finder (AHF, Gill et al.~2004 and Knollmann \& Knebe 2009) as our dark matter halo (DMH) identification algorithm which uses an adaptive mesh to look for bound particle systems.  We selected halos with a minimum number of particles $N_p=100$, which corresponds in our simulations to halos with masses $M_{\rm min}\approx 6\times 10^{10}$\hmsun, in order to have well defined halos not much subject to numerical noise. The output of the AHF code provides, among other things, the viral mass and radius of halos and subhalos.

%%%%%%%%%%%%%%%%%
\begin{figure}
\centering
  \includegraphics[width=0.7\columnwidth]{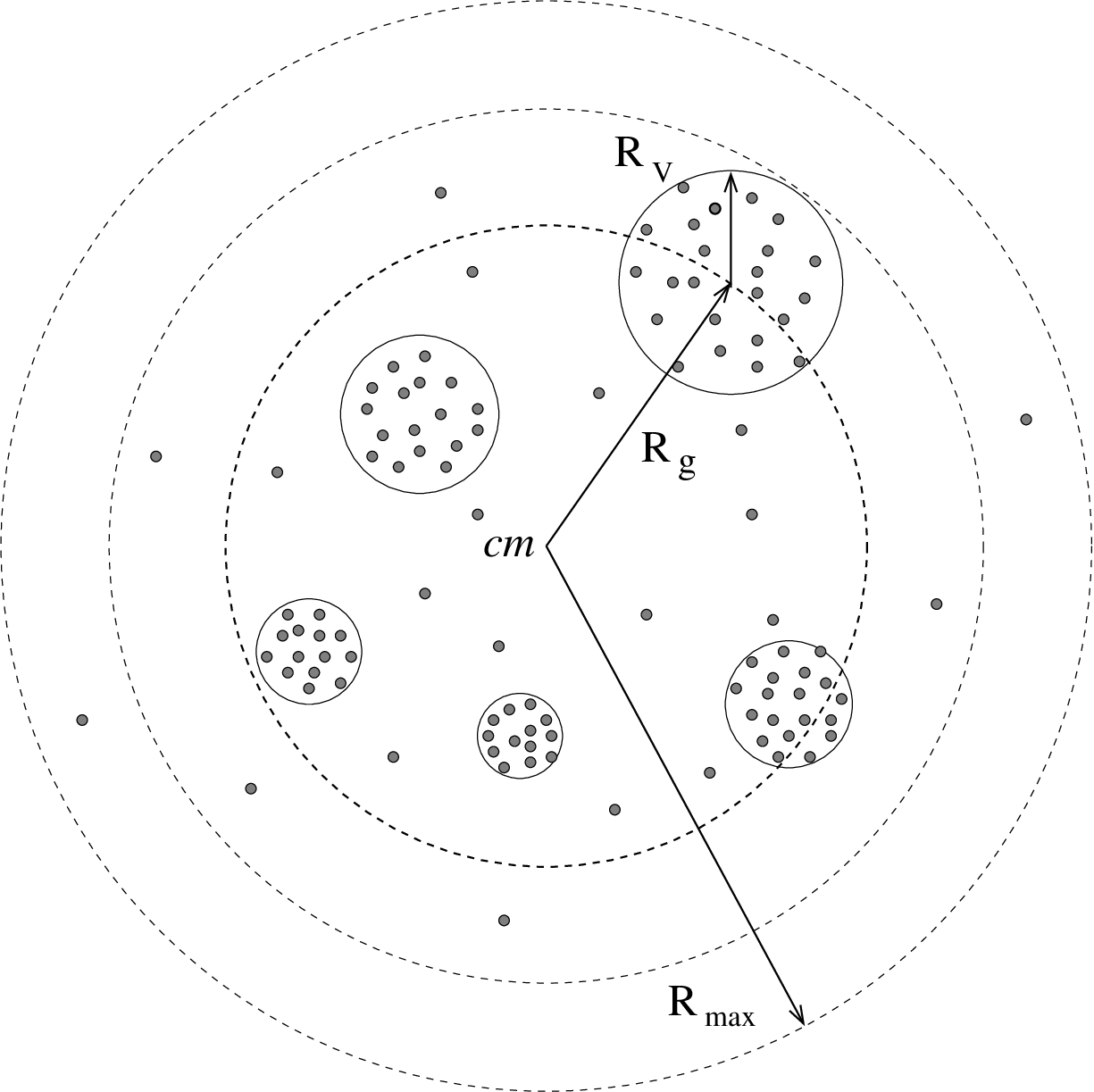}
  \caption{Schematic diagram of different quantities used to determine our physical groups. Matter not bound to individual halos, of virial radius $R_{\rm V}$, is considered to belong to the intragroup medium, if it is also located within $R_{\rm max}\!=\! 1\,h^{-1}$Mpc from the center-of-mass ($cm$) of the group.  The group radius, $R_{\rm g}$, is determined by the radius of an imaginary  three-dimensional sphere centered at its center-of-mass and extended up to the center of the furthest galaxy-like halo.}
  \label{fig:group}
\end{figure}
%%%%%%%%%%%%%%%%%%%%%

%%%%%%%%%%%%%%
\subsubsection{Groups}\label{sec:groups}
%%%%%%%%%%%%%%

As noted in $\S$\ref{sec:intro} it is not the purpose here to make a mock catalog resembling small galaxy groups, loose or compact, or to make a direct comparison with observations. Our objective is to determine {\sl physically} bound small groups of halos in our $\Lambda$CDM cosmological simulations; these groups nonetheless may resemble small galaxy groups in regard to their intragroup dark matter distribution. In order to do the previous  we define clear physical quantities in our search algorithm. We proceeded as follows to determine a group of halos probably hosting ``normal'' galaxies. 

First, taking the galaxies of our Local Group as being typical of  a small galaxy group environment, we determine a halo mass that can be associated with a galaxy like M33. This galaxy is taken as our fiducial total lower mass for a ``normal'' galaxy.  Using the monotonic mass-luminosity relation of Vale \& Ostriker (2004)  with an absolute magnitude $M_V\!=\!-18.9$ for M33 (Mo, van den Bosch \& White 2010), an  estimate of total mass of $M_{\rm min}\approx 10^{11}$\hmsun\/ is obtained, which is consistent with the value used by Berlind~et~al.~(2006) from Halo Occupation Distribution  fits to the SDSS two-point correlation function of galaxies for such luminosity. Hence, we take $M_{\rm min}$ as the lowest mass of the dark halo of a normal galaxy to be considered as part of a group of halos. The upper mass was set to be about twice that of the Milky Way, with $M_{\rm max} \approx 5\times 10^{12}$\hmsun\/. 
Hence, in our simulations we used halos belonging to the a mass range $M\in [M_{\rm min},M_{\rm max}]$ as a criterion to determining membership in a small group of halos. In other words, we excluded smaller subhalos from defining a group of galaxy-size halos in our approach, as well as very massive single halos ($M\gta M_{\rm max}$) that are not found in small galaxy groups; we do not consider here fossil groups that may host a cD-type galaxy.

Secondly, we used a simple search algorithm to determine our physical groups of halos  at $z=0$. This algorithm required that the number of galaxy-size halos $N_{\rm h}$ to be $N_{\rm h}\in [4,10]$ and within a physical radius of $R_{\rm max}=1\,h^{-1}$Mpc from the center of mass of the tentative members, and that no other normal galaxy-size halo be within $R_{\rm n}\! =\! 1.25\,R_{\rm max}$. The radius  $R_{\rm max}$ chosen corresponds to about the turn-around radius (Gunn \& Gott~1972) of a mass of $\approx \! 10^{13}$M$_\odot$, and $R_{\rm n}$ is set only to provide a clear physical isolation criteria from other possible nearby bound structures; see Figure~\ref{fig:group}. 
We applied this group-search algorithm in a marching way to all dark halos that have $M\in [M_{\rm min},M_{\rm max}]$ in the simulations. This procedure generated a set of group candidates with radius $R_{\rm g}$; measured from the center-of-mass of the galaxy-size halos to the center of the  outermost one.

In order to differentiate several degrees of compactness found in the dark  groups, we refer to those groups with a spherical radius $R_{\rm g}<250\,h^{-1}\,$kpc as  compact associations (CAs, or compact groups for simplicity), intermediate associations (IAs) to those systems with $R_{\rm g}\in (250,500)\,h^{-1}\,$kpc and loose associations (LAs, or loose groups) for groups with $R_{\rm g}\in [500,1000]\,h^{-1}\,$kpc.

 In our previous set of group candidates their isolation degree in the cosmological simulations was needed to be further adjusted to have a more clean sample of isolated groups. This was specially required for LAs since some of them had larger structures within $\approx 1\,h^{-1}$Mpc of its outermost halo. All of the CAs satisfied an isolation criteria similar to that of Hickson of not having a galaxy-size dark halo within $3 R_{\rm g}$, since for CAs its maximum group radius satisfies by definition $3 R_{\rm g}< R_{\rm max}$, so no further modifications in the algorithm were required. For both IAs and LAs we imposed a further restriction that no other dark halo would be within $R_{\rm g}+R_{\rm max}$ from their center. This allowed us to have a well isolated sample of IAs and LAs.

Finally, we checked that all groups identified with the above procedure were actually bound systems by estimating their kinetic and potential energy  in an approximate manner as if the halos were point particles,  and the intragroup matter was negligible. The kinetic energy of the group was computed from
\begin{equation}
T = \frac{1}{2M} \sum_{i<j} M_i M_j ( \mathbf{V}_i - \mathbf{V}_j )^2 \,,
\end{equation}
with $M$ being the total mass of the group of virialized halos, $M_i$ that of the $i$-th halo, and $\mathbf{V}_i$ the corresponding velocity. The potential energy was calculated as
\begin{equation}
U = -G \sum_{i<j} \frac{M_i  M_j}{R_{ij}} \,,
\end{equation}
with $R_{ij}$ the physical separation between two halos. We considered the group to be bound if $T<|U|$.  A similar approximate approach was used by Niemi et~al.~(2007) to discriminate bound from unbound groups of halos.  All group of halos determined in the previous step to this were bound. The average virial ratio of all of our groups was  $\langle 2T/|U| \rangle = 0.14 \pm 0.33$, while for CAs  was $0.03\pm 0.08$. As shown in Section~3, all CAs found were in a collapsing state in the cosmological simulations; consistent with their low average virial ratio.

If one would be interested in comparing with observations, a way to assign luminosity to the dark halos would be required in order to compare with observational catalogues (e.g. e.g. Casagrande \& Diaferio 2006, McConnachie et al.~2008, D\'{\i}az-Gim\'enez \& Mamon 2007, and  Niemi et al.~2007). However, the latter is out of the scope of the present work that does not intend such comparison with observations.

%%%%%%%%%%%%%%%%%
\begin{figure*}[t]
\centering
  \includegraphics[width=0.95\columnwidth]{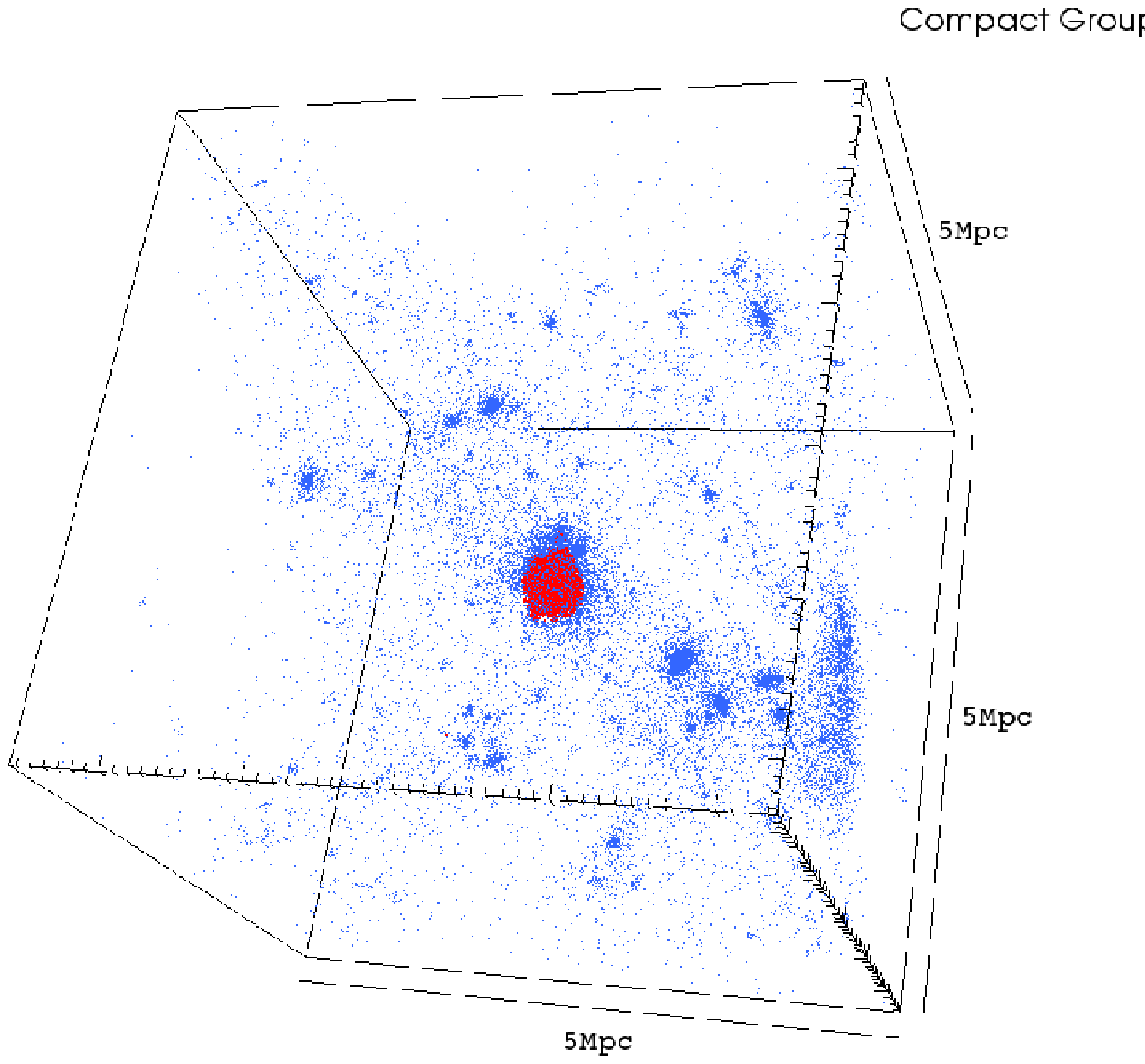}
  \hspace{0.5cm}
  \includegraphics[width=0.95\columnwidth]{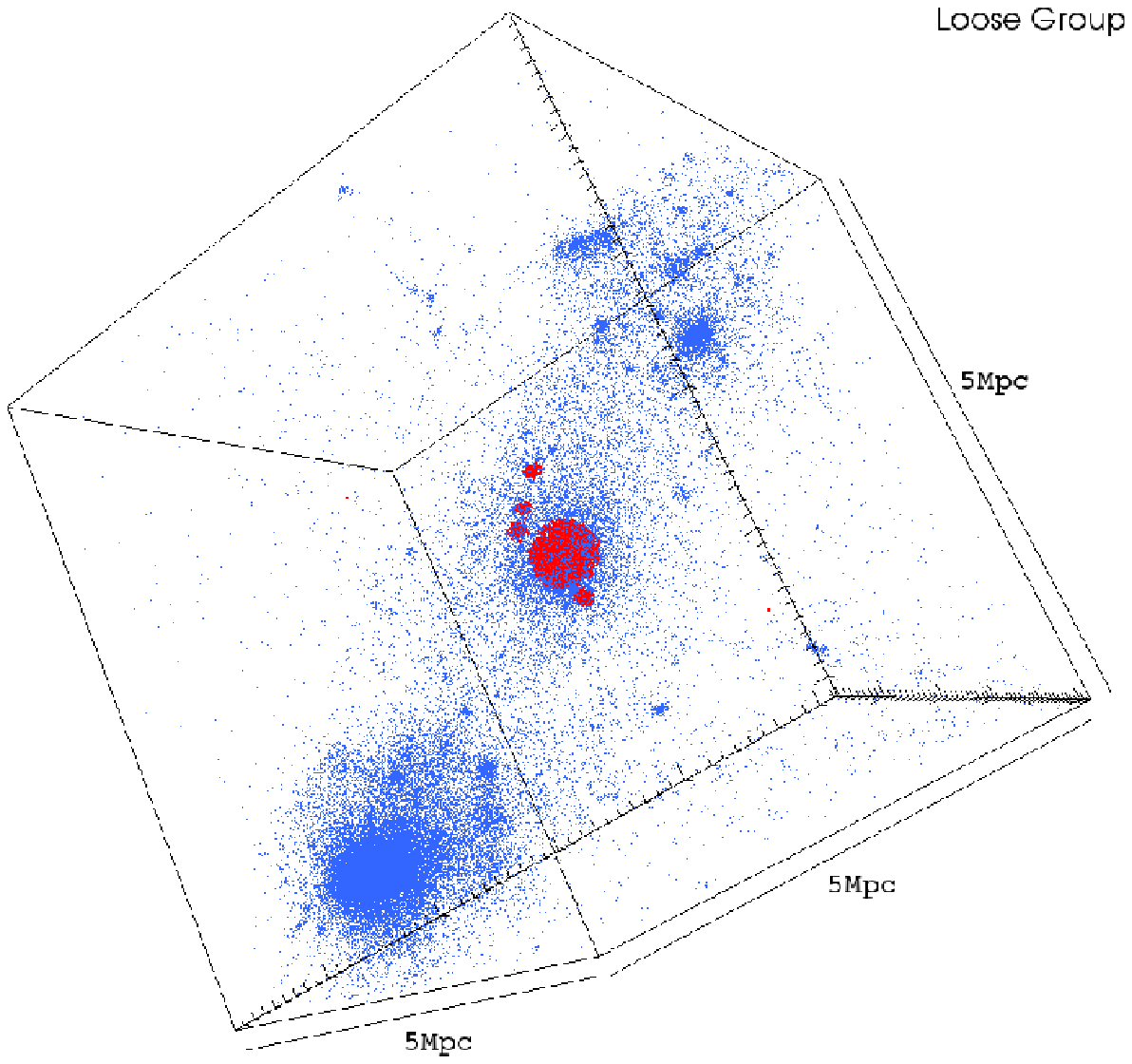}
  \caption{An example of a compact ({\it left}) and loose ({\it right}) group  association of halos. The red dots (see the online version of the manuscript) indicate particles belonging to halos that satisfy our group criteria, and blue dots refer to particles belonging to the intragroup medium; other structures not belonging to such groups are seen. The two small dark halo groups are physically bound objects, not projected systems.}
  \label{fig:groups}
\end{figure*}
%%%%%%%%%%%%%%%%%%%%%

It is known that different identification criteria lead to a different number of galaxy associations found in a simulation or in the sky (e.g. Duarte \& Mamon~2014).
In the our method we selected groups of halos in a way we consider mimic in a simple way the procedure used to determine galaxy groups  (e.g. Lee et al.~2004). Namely, without using any assumption aside of considering galaxy-like mass halos, we look for associations of dark halos within a certain spatial region. Then we verified that they are truly physical bounded groups and explore their properties. Other recent approaches (e.g. Berlind et al.~2006, Yang~et~al.~2007, Dom\'{\i}nguez-Romero~et~al.~2012) essentially go the other way around, they  look for dark halos of a particular mass range ($\approx 10^{13}$\hmsun) and their properties and content (subhalos) are studied. We discuss  in a brief way these differences with our approach later in  the paper and how it relates to our results.

Finally, to quantify the distribution of intra-group dark matter we removed all particles not associated with individual halos, as signaled by the AHF code, and measured its amount and determined its distribution with respect to the group center-of-mass. The total intragroup mass was determined both inside the group radius $R_{\rm g}$ and the estimated turn-around radius $R_{\rm max}$. We traced the evolution of all particles in CAs from $z\!=\!0.5$ to $z\!=\!0$ and all remained within the $R_{\rm max}$ radius.  We estimated the mass density profile,  in particular to determine its inner slope,  assuming spherical symmetry for the IG dark matter, and by stacking about the same number of IG particles for CAs and LAs.

%%%%%%%%%%%%%%%%%%%%%%%%%%%%%%%%%%%%%%%%%%%%%%%%%%%%%%%%%%%%%%%%%%%%%
%%%%%%%%%%%%%%%%%%%%%%%%%%%%%%%%%%%%%%%%%%%%%%%%%%%%%%%%%%%%%%%%%%%%%
\section{Results}
\label{sec:results}
%%%%%%%%%%%%%%%%%%%%%%%%%%%%%%%%%%%%%%%%%%%%%%%%%%%%%%%%%%%%%%%%%%%%%%%%%
%%%%%%%%%%%%%%%%%%%%%%%%%%%%%%%%%%%%%%%%%%%%%%%%%%%%%%%%%%%%%%%%%%%%%%%%%

In Figure~\ref{fig:groups} we show the location of two small dark groups found in one of our cosmological simulations. In red we show the particles associated to halos of galaxies belonging to a group according to our selection criteria above, and in blue the matter not associated to the DMHs of the group. Shown in Figure~\ref{fig:groups} are only compact ({\sl left}) and loose ({\sl right}) association of halos. A visual inspection of our simulations showed that most of our groups are found in filaments of the large-scale structure (e.g. Hernquist et al.~1995), although some of them are near larger cluster-like structures at the nodes of the cosmic web. None was found at what might be called voids. All results are  consistent with the general trends of observations of small galaxy groups (e.g. God{\l}owski \& Flin~2010, Mendel et al.~2011).  In Table~1 the average values and standard deviation of the total mass (halos and IG matter), radius of the group, three-dimensional velocity dispersion and dimensionless crossing time are reported for our associations.

\begin{table*}[t]
\caption{Global properties of Dark Halos Groups \label{fig:tab1}}
\centering
\begin{tabular}{cllll}\toprule
   & $M$ & $R_{\rm g}$ & $\sigma$  & $t_c/t_H$ \\ 
   & $10^{12}h^{-1}\,$M$_\odot$ & $h^{-1}\,$kpc & km/s  & \\ \midrule
CAs  &    $9.23 \pm 2.23$    & $211.2 \pm 35.7$  & $296.0 \pm 59.9$  & $0.038 \pm 0.010 $     \\
IAs  &   $7.46 \pm 4.88$          & $419.2 \pm 61.9$  & $296.3 \pm 142.2$  & $0.088 \pm 0.041 $     \\
LAs  &     $8.01 \pm 4.77$     & $827.5 \pm 121.9$  & $263.2 \pm 150.2$  & $1.349 \pm 18.032 $     \\ \bottomrule
\end{tabular}
\end{table*}

In our five cosmological simulations we found at $z=0$ a total number of  14 objects classified as CAs, a set of 64 IAs, and a total of 661 LAs.
 In particular the number of CAs in average per simulation is $\langle N_{CA} \rangle \approx 3$ in our box of volume $(100\,h^{-1}{\rm Mpc})^3$. This   number of CAs  appears to be rather low  when comparing with results of other authors. Scaling these numbers to boxes as the one in the {\sl Millennium Simulation}  ({\sl MS}, Springel et al.~2005) of side length  $L=500\,h^{-1}\,$Mpc,  such that we multiply our numbers by a volume correction factor of $5^3$, we are at  about $1/3$ of CAs found for example by McConnachie et al. (2008) who found $\approx 1200$ groups in the {\sl MS}; albeit using another search algorithm, the numbers are  significantly different.

 We tested if the use of a higher value of $\sigma_8=0.91$ as used in the {\sl MS}, in comparison to the value used here $\sigma_8=0.82$, may have lead to more structures at the scale of groups and, hence, to obtain perhaps a better consistency with other works that use the {\sl MS} results.
We made two additional  cosmological simulations with the same cosmological parameters as the {\sl MS} (e.g. $\Omega_m=0.3$, $\Omega_\Lambda=0.7$ and $\sigma_8=0.91$) but in a box of length $L=100\,\textrm{Mpc/h}$ and with $512^3$ particles.

%%%%%%%%%%%%%%%%%
\begin{figure*}[t]
\centering
  \includegraphics[width=0.9\columnwidth]{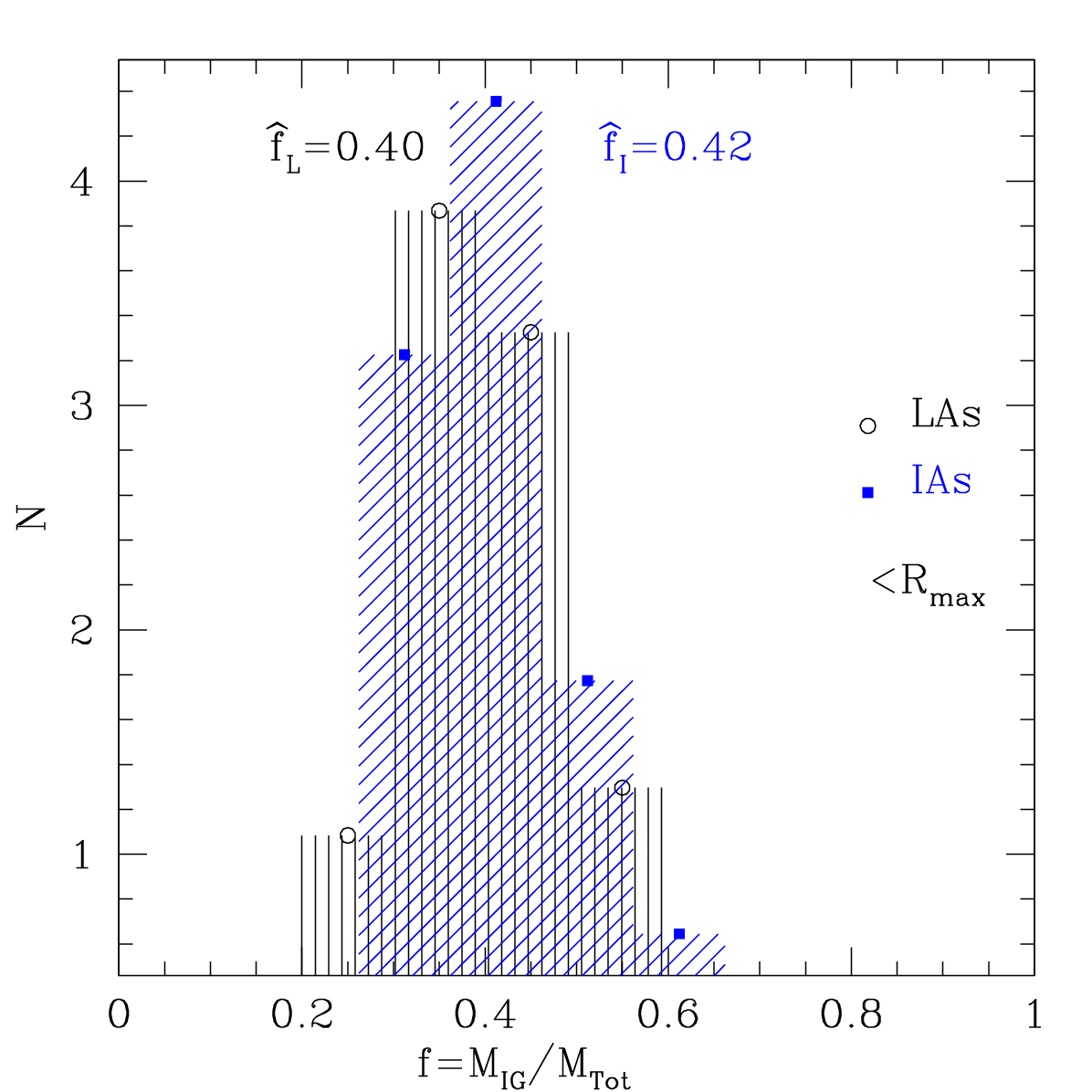}
  \includegraphics[width=0.9\columnwidth]{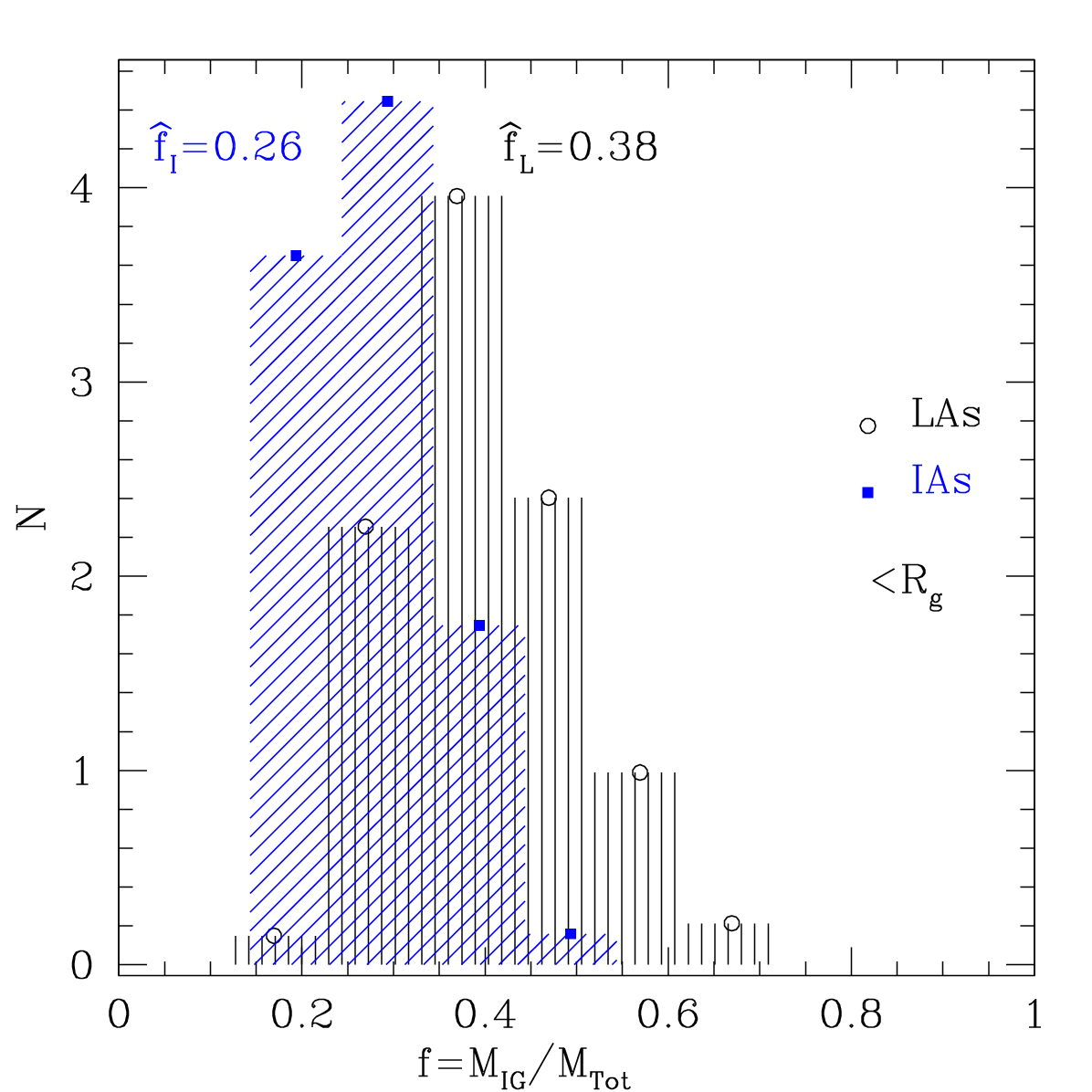}
  \caption{Frequency distribution of intragroup dark matter to total mass in  our small group-like objects, both within $R_{\rm max}$ ({\it left}) and the radius $R_{\rm g}$ ({\it right}). Median values are indicated for loose and intermediate associations. For compact associations we obtained an average value of $\langle f \rangle_{\rm C}= 0.41$ in the first case and $\langle f \rangle_{\rm C}= 0.20$ in the second one. }
  \label{fig:histo}
\end{figure*}
%%%%%%%%%%%%%%%%%%%%%

Performing the same analysis  to identify CAs as in our  two  simulations above, we found an average of $\langle N_{CA} \rangle_{MS} =10$.
 This increase of CAs in the {\sl MS}-like simulations is consistent with the expected result when higher $\sigma_8$ is used in a cosmological model, albeit we are dealing here with a small number statistics.  
After correcting by the volume factor indicated above,  the number of CAs are similar as those of other authors for these type of groups that use similar selection criteria,  but including luminosity related properties  (e.g. Casagrande \& Diaferio 2006, McConnachie et al. 2008, D\'{\i}az-Gim\'enez \& Mamon 2007, Niemi et al.~2007).  Other  differences of lower order may arise, for instance, due to using a friends-of-friends criteria (e.g.  McConnachie et al. 2008) when fixing the size of the linking length $l$ of the ``friendship'' (e.g. Duarte \& Mamon~2014), or with lowering the observational threshold ``magnitude'' of detection in groups in mock catalogs.

\subsection{Amount of IG dark matter}
%%%%%%%%%%%%%%%%%%%%%%%%%%%%%

As indicated in $\S$\ref{sec:methods} we identify dark matter particles associated with an intragroup environment as those not physically bound to the  haloes identified with the AHF code; first all particles within a $R_{\rm max} \!=\!1\,h^{-1}$Mpc radius and then within $R_{\rm g}$. In  Figure~\ref{fig:histo} we show the frequency of the ratio of IG dark matter to total group mass, $f$, for our loose associations as well as for the intermediate ones, using both group identification criteria. The average total, IG and the matter in halos,  mass obtained for all three type of groups found was $\langle M \rangle \approx 8 \times 10^{12} h^{-1} {\rm M}_\odot$.

The median values of such ratios, within $R_{\rm max}$, is ${\hat f}_{\rm L}=0.40$ and ${\hat f}_{\rm I}=0.42$ for our loose and intermediate associations, respectively. For compact associations we find and average value of $\langle f \rangle_{\rm C}= 0.41$, but not shown in Figure~\ref{fig:histo} as a histogram since we only have a few points. 
In general, the amount of IG dark matter tends to  be less than $\approx 50$\% of that of the whole bound system, with a median of $\approx 40$\%  irrespective of the configuration of the group if the size of the group is taken to be $R_{\rm max}$.

When we counted only matter within the group radius $R_{\rm g}$ for CAs  we obtained $\langle f \rangle_{\rm C}= 0.20$, and for IAs and LAs we find the corresponding fractions to be ${\hat f}_{\rm I}=0.26$  and ${\hat f}_{\rm  L}=0.38$, respectively.  As noted the difference between the two ways to determine the size of the group, in regards to the intragroup dark matter fraction, tends to agree only for loose associations and becomes more  significantly different for the compact groups.

It is to note that within the group radius  $R_{\rm g}$, specially for compact configurations, a lot of dark matter particles find their way into bound structures thus reducing the intragroup medium; since the latter is defined precisely by particles not bound to any halo of the member galaxies. This behavior is noted also in the density profiles computed in the next section. Some matter may also be associated to smaller subhalo type structures, but we do not distinguish here from dark subhalo particles and intragroup particles.

%%%%%%%%%%%%%%%%%
\begin{figure}[t]
\centering
  \includegraphics[width=0.9\columnwidth]{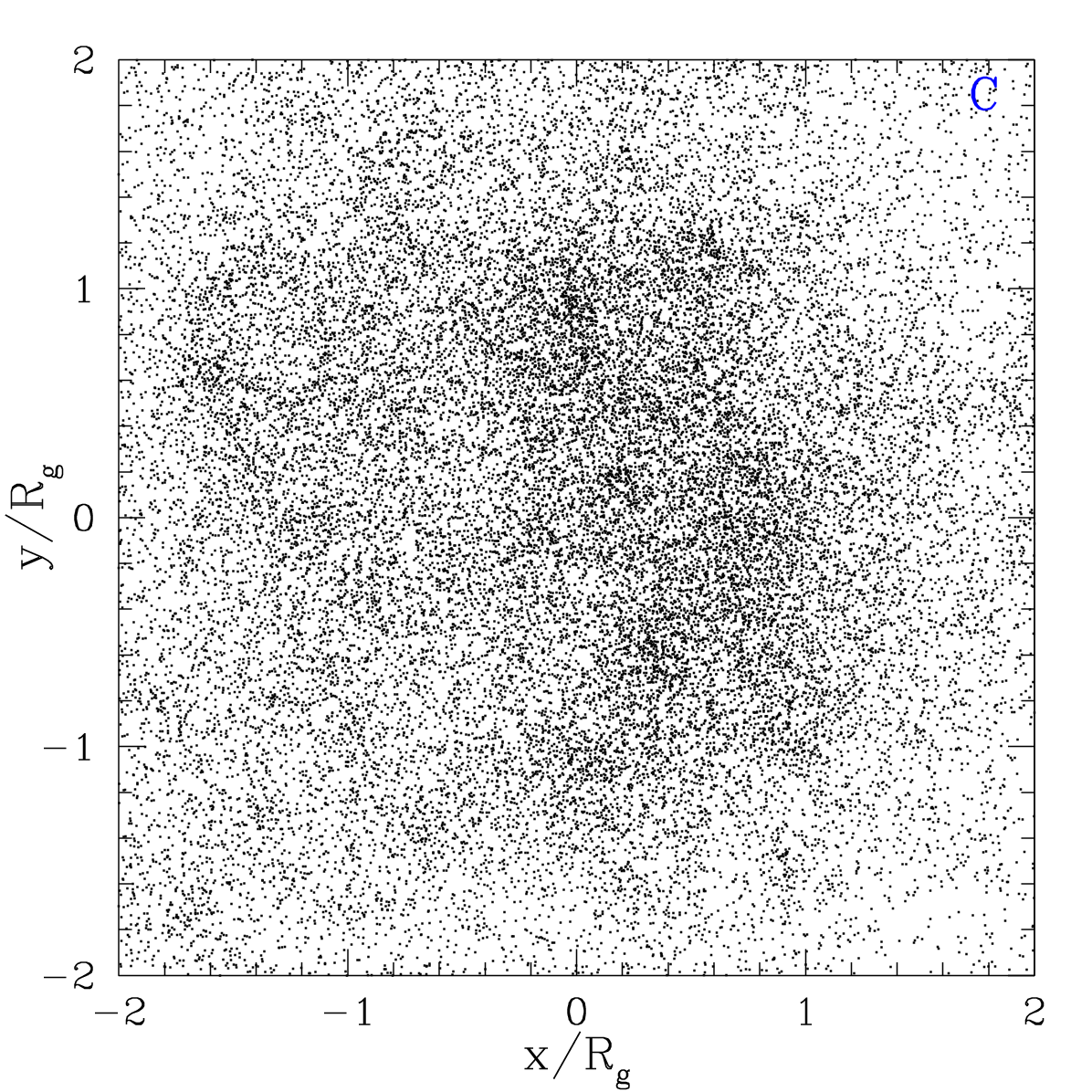}
  \caption{Stacked distribution of IG dark matter of small groups of dark halos that have a compact configuration.  The total number of particles shown are 39,231 and the thickness of the box is $4 R_{\rm g}$.  }
  \label{fig:cgroupsDM}
\end{figure}
%%%%%%%%%%%%%%%%%%%%%

%%%%%%%%%%%%%%%%%
\begin{figure}[t]
\centering
  \includegraphics[width=0.9\columnwidth]{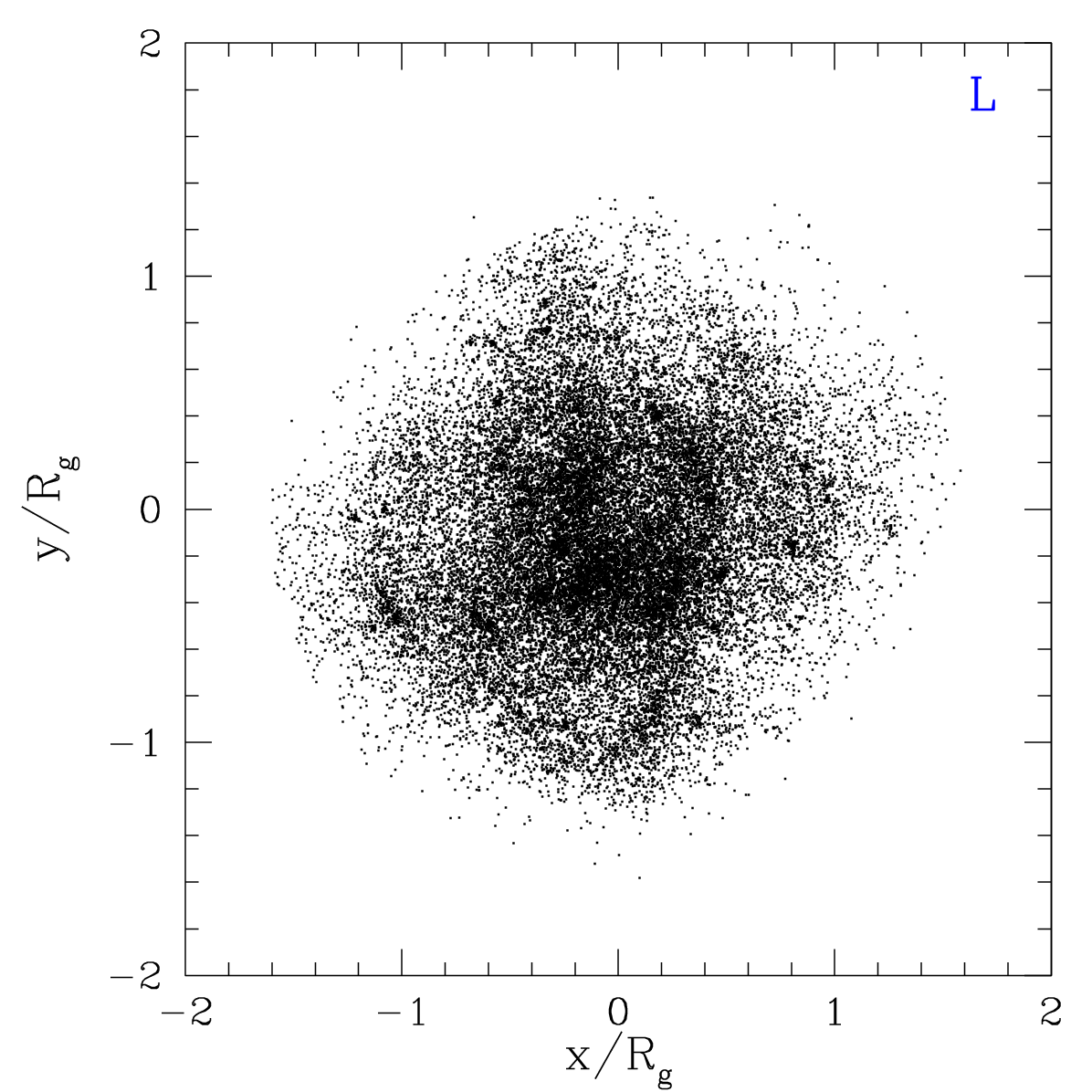}
  \caption{As in Fig.~\ref{fig:cgroupsDM} but for loose groups.  Plotted are just 34,112 IG dark particles to aid in viewing some residual structures in the dark matter; i.e. not considered to be bound halos due to our selection criteria.}
  \label{fig:lgroupsDM}
\end{figure}
%%%%%%%%%%%%%%%%%%%%%

\subsection{IG dark matter profiles}
%%%%%%%%%%%%%%%%%%%%%%%%%%%%%

In Figure~\ref{fig:cgroupsDM} we show the stacked distribution of  about 39,000 IG dark matter particles of compact associations found in our simulations at $z\!=\!0$ according to our selection criteria. A similar plot but for loose groups is shown in Figure~\ref{fig:lgroupsDM};  here $\approx 35,000$ particles are shown. The scale of both figures is the same in terms of the group radius of the associations. The center of mass of all  stacked groups coincide. We should mention that no galaxy-size dark halo was found to reside at the center of mass of a group in our search algorithm.

Both Figures~\ref{fig:cgroupsDM} and~\ref{fig:lgroupsDM} show no indications of a central concentration of intragroup dark matter and show a behavior more akin to an homogeneous distribution;  the same behavior was observed for the intermediate associations. In these plots we have taken out all the particles belonging to halos with $N_{\rm p}\ge 100$ out of the accounting, but smaller concentrations appear in them.

%%%%%%%%%%%%%%%%%
\begin{figure*}[t]
\centering
  \includegraphics[width=0.9\columnwidth]{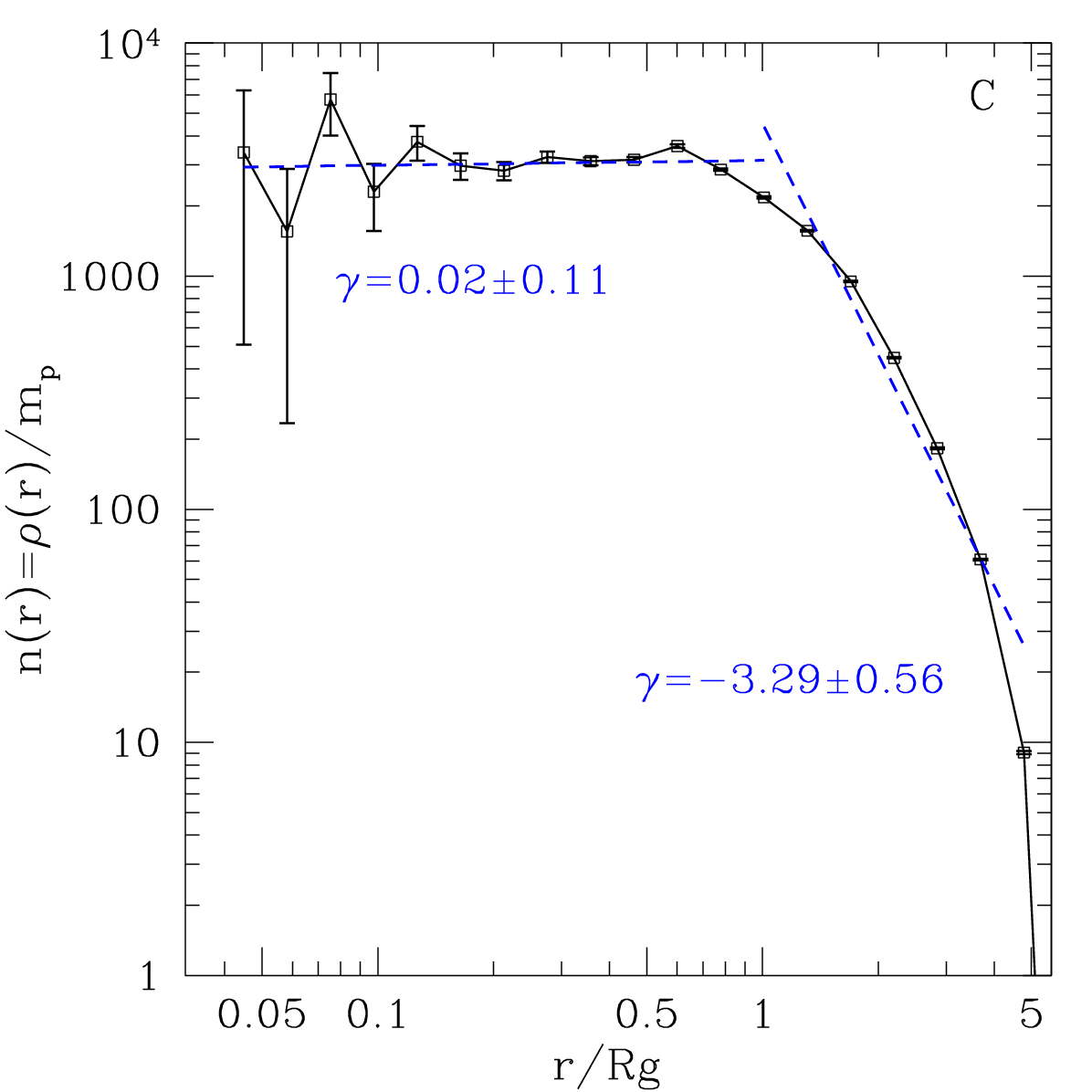}
  \hspace{10pt}
  \includegraphics[width=0.9\columnwidth]{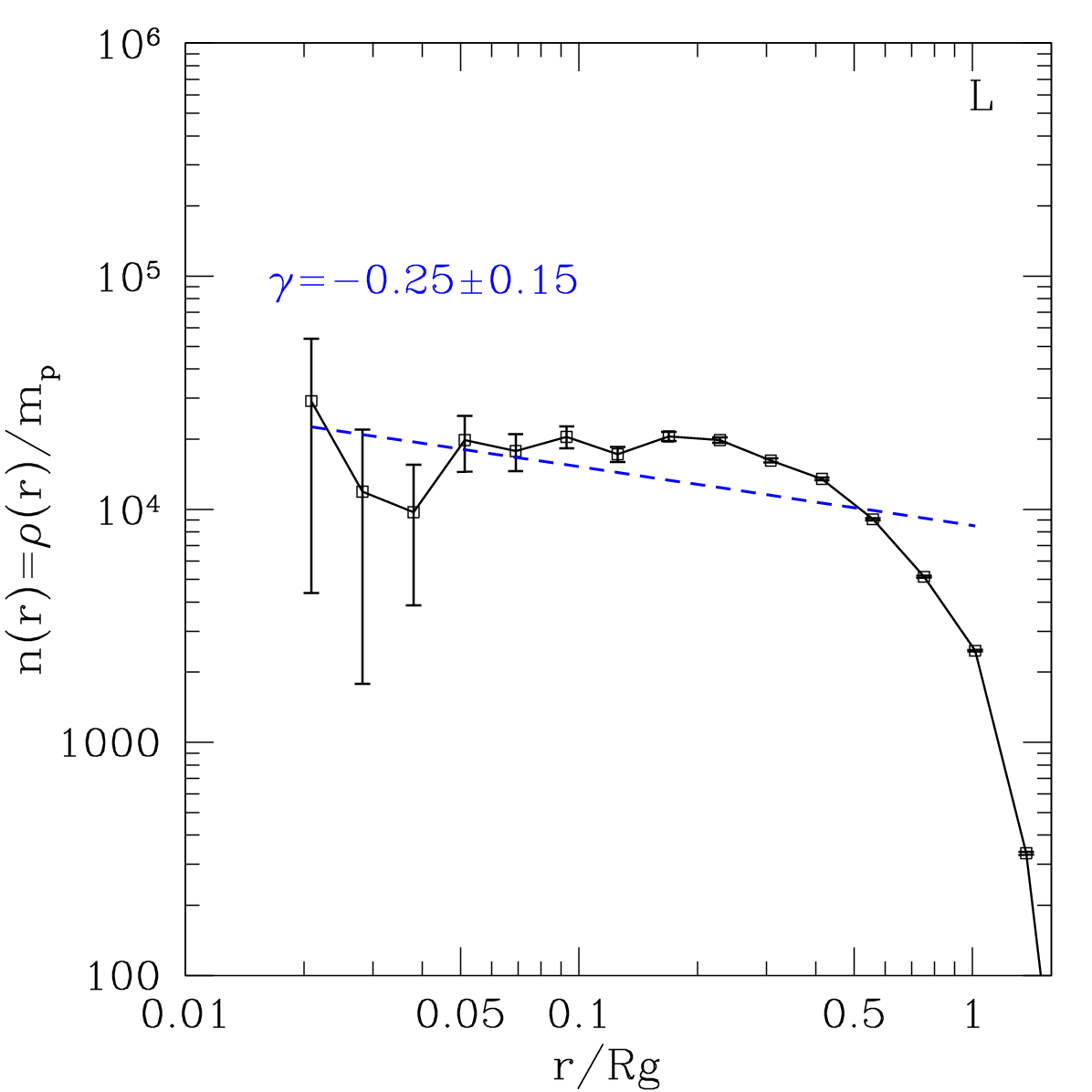}
  \vspace{-5pt}
  \caption{IG dark matter density profile of compact  (CAs, {\it left}) and loose associations (LAs, {\it Right}). Compact groups  of galaxy-size dark halos show an inner power-slope of $\gamma=0.02\pm 0.11$ at $r\le R_{\rm g}$,  with a  rapid density decay afterwards. Loose groups up to  their $R_{\rm g}$ yield a $\gamma=-0.25\pm0.15$ slope. }
  \label{fig:groupsP}
\end{figure*}
%%%%%%%%%%%%%%%%%%%%%

To quantify the degree of concentration of the IG dark matter we computed the  spherically-averaged density profile $\rho(r)$ of all particles belonging to compact, intermediate and loose associations;  the profile is centered on the center-of-mass of the group as defined earlier.  All coordinates of particles have been scaled by the group radius to which they belong when computing $\rho(r)$.

 In Figure~\ref{fig:groupsP}({\it left}) the mass density profile of the IG-DM is shown for compact associations.  A power-slope fit, $\rho \propto r^{\gamma}$,  was made yielding a value of $\gamma=0.02\pm 0.11$ for the inner part  $r \le R_{\rm g}$,  and an external slope ($r\in (R_{\rm g},5 R_{\rm g}]$)  of  $\gamma=-3.29\pm 0.56$ was obtained. Errors in the fits to the slopes of the profiles were estimated in all cases by a bootstrap method  (Efron \& Tibshirani~1993).  Errorbars at the data points are Poisson errors.

In Figure~\ref{fig:groupsP}({\it right}) we show the IG dark matter profile for those groups identified as loose. An internal power-slope fit yields $\gamma=-0.25\pm0.15$; no fit is made after $R_{\rm g}$ since for these  groups $R_{\rm g} \approx R_{\rm max}$ where the density falls rather sharply after $R_{\rm g}$.

 The profiles shown in Figure~\ref{fig:groupsP} have a minimum starting radius at about the scale of the softening radius ($\varepsilon =  20 h^{-1}$kpc), in terms of the average group radius for the halo associations shown. In order to explore the effect of the softening radius on the inner slope of the halo associations, we made two additional cosmological simulations but with $\varepsilon = 1 h^{-1}$kpc. For 6 CAs identified in these new simulations, after stacking them, we obtained an inner slope of $\gamma =2.09 \pm 0.18$ and an external one of $\gamma = -4.01 \pm 0.37$, and for the 273 LAs an inner slope of $\gamma =0.30 \pm 0.13$ results. The deficiency of IG dark matter in this sample of CAs, with inner slope $\gamma \approx 2$, is also noticed in several systems of the first set of simulations. The two old  simulations similar to the new ones, both using the same initial seed to construct the initial conditions, yield an inner slope of $\gamma = 1.84\pm 0.13$ for these 6 groups. The two inner slopes for the CAs, for the two different softenings considered here, are consistent with each other; the same is observed for the outer slopes.
However, the reported value above  using $\varepsilon =  20 h^{-1}$kpc  results from stacking more associations and can thus be considered an average behavior in a typical systems of this kind. The previous situation is indicative  nonetheless of the complexity of the inner distribution of dark matter in compact associations of galaxy-size dark halos.  

It follows, however, from the above results that the IG dark matter does {\it not} tend  to be  cuspy in any kind of the halo associations found here, and also that it does not dominate in mass.

\subsection{Evolutionary trends}
%%%%%%%%%%%%%%%%%%%%%%%%%%%%%

\begin{figure*}
\centering
  \includegraphics[width=0.65\columnwidth]{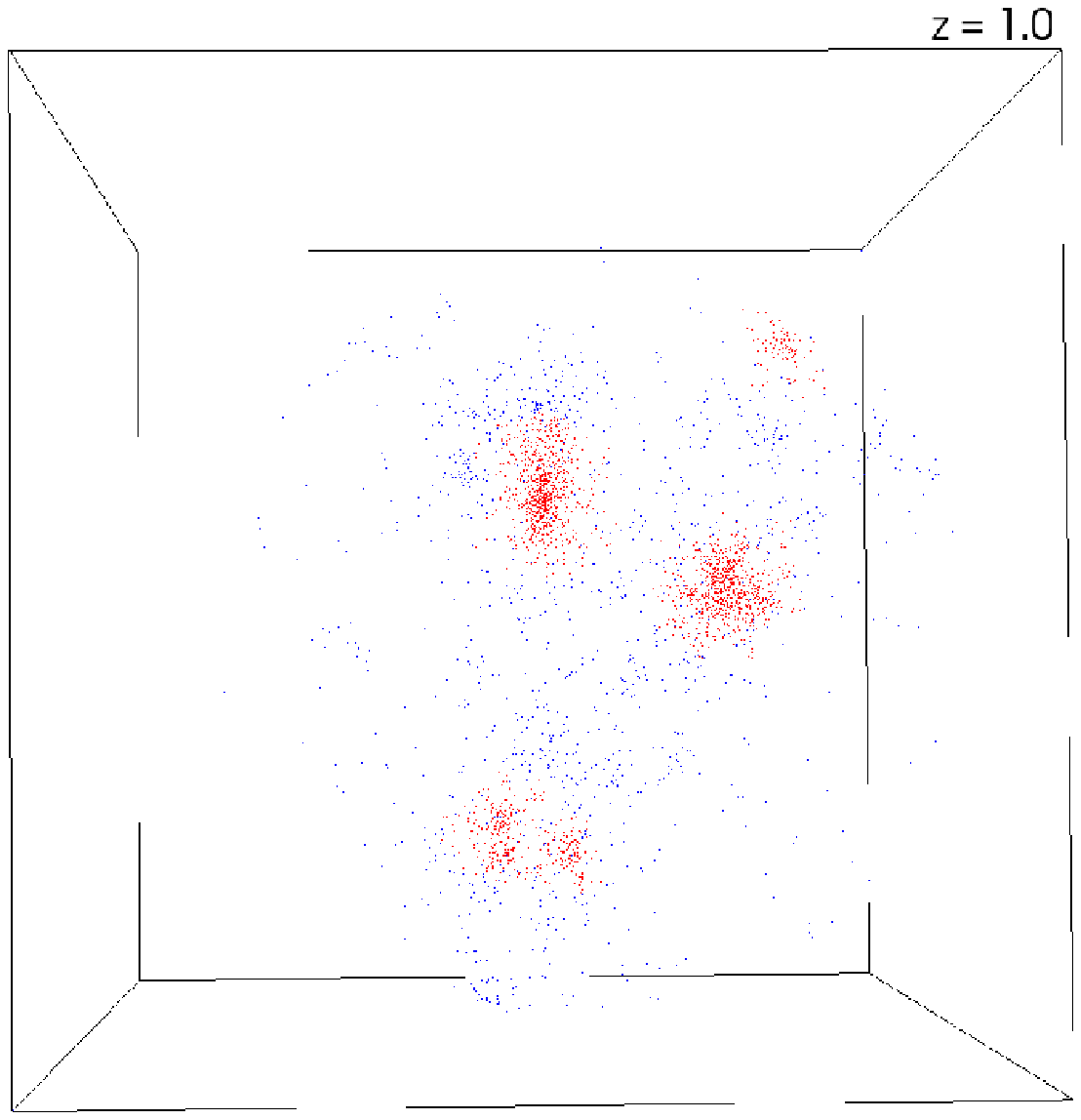}
  \includegraphics[width=0.65\columnwidth]{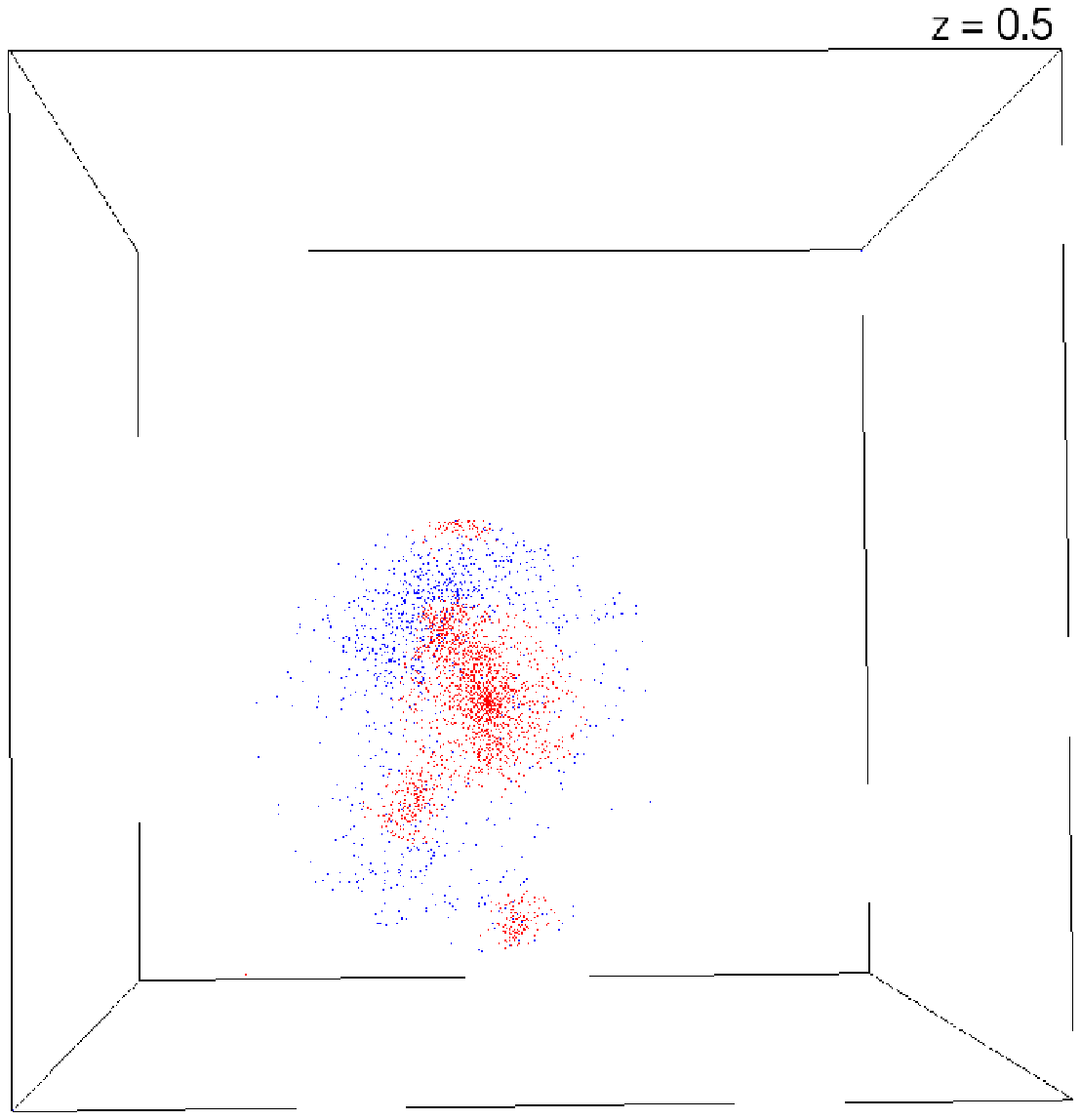}
  \includegraphics[width=0.65\columnwidth]{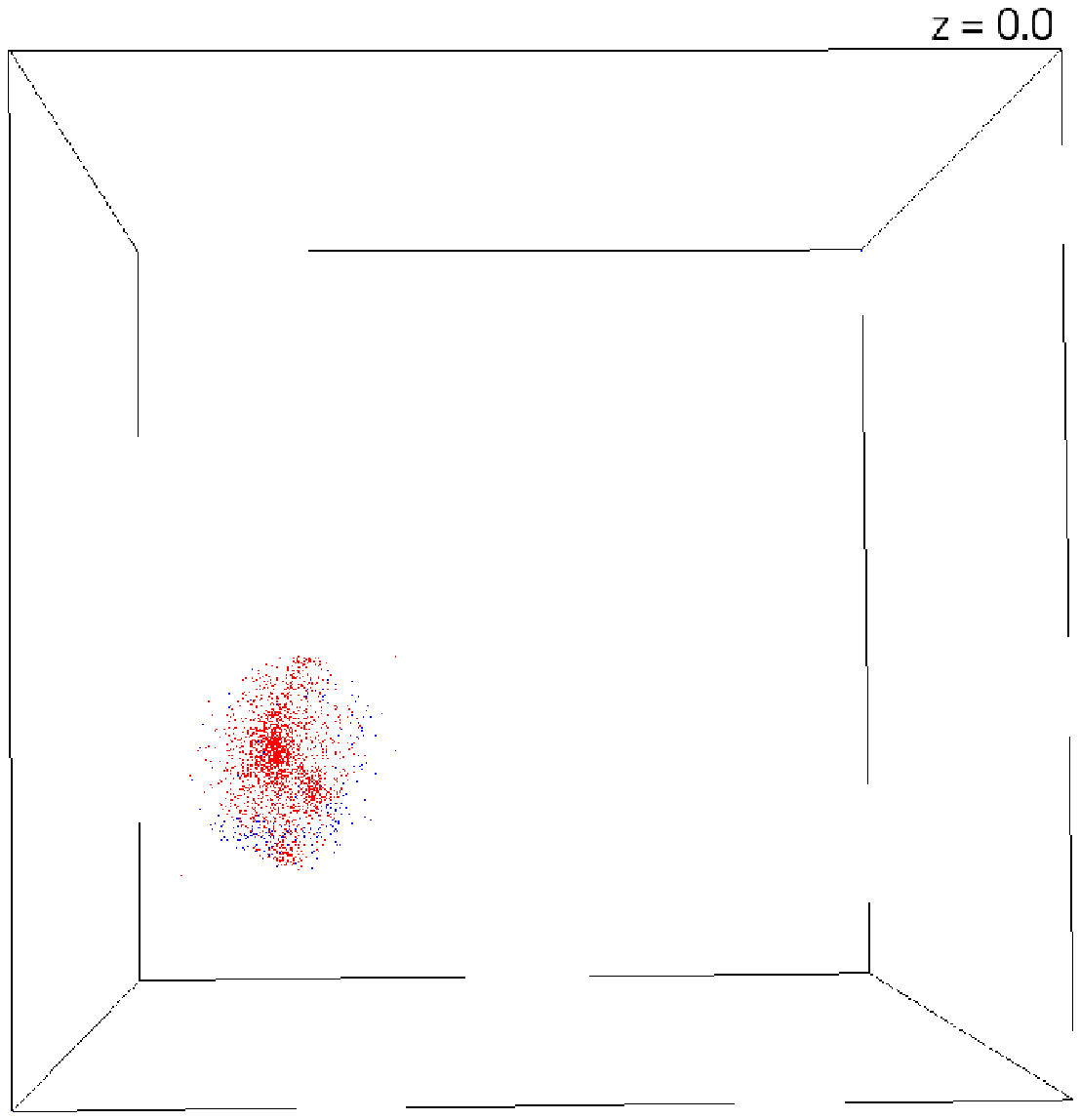}
  \caption{Evolutionary sequence of a particular compact group, at $z=1.0$, $z=0.5$ and $z=0$. The box has $2\,h^{-1}\,$Mpc on each side. Red points correspond to particles associated with galaxy haloes while blue ones to intragroup dark matter. }
  \label{fig:groupsEvol}
\end{figure*}

\begin{figure}[t]
\centering
  \includegraphics[width=0.85\columnwidth]{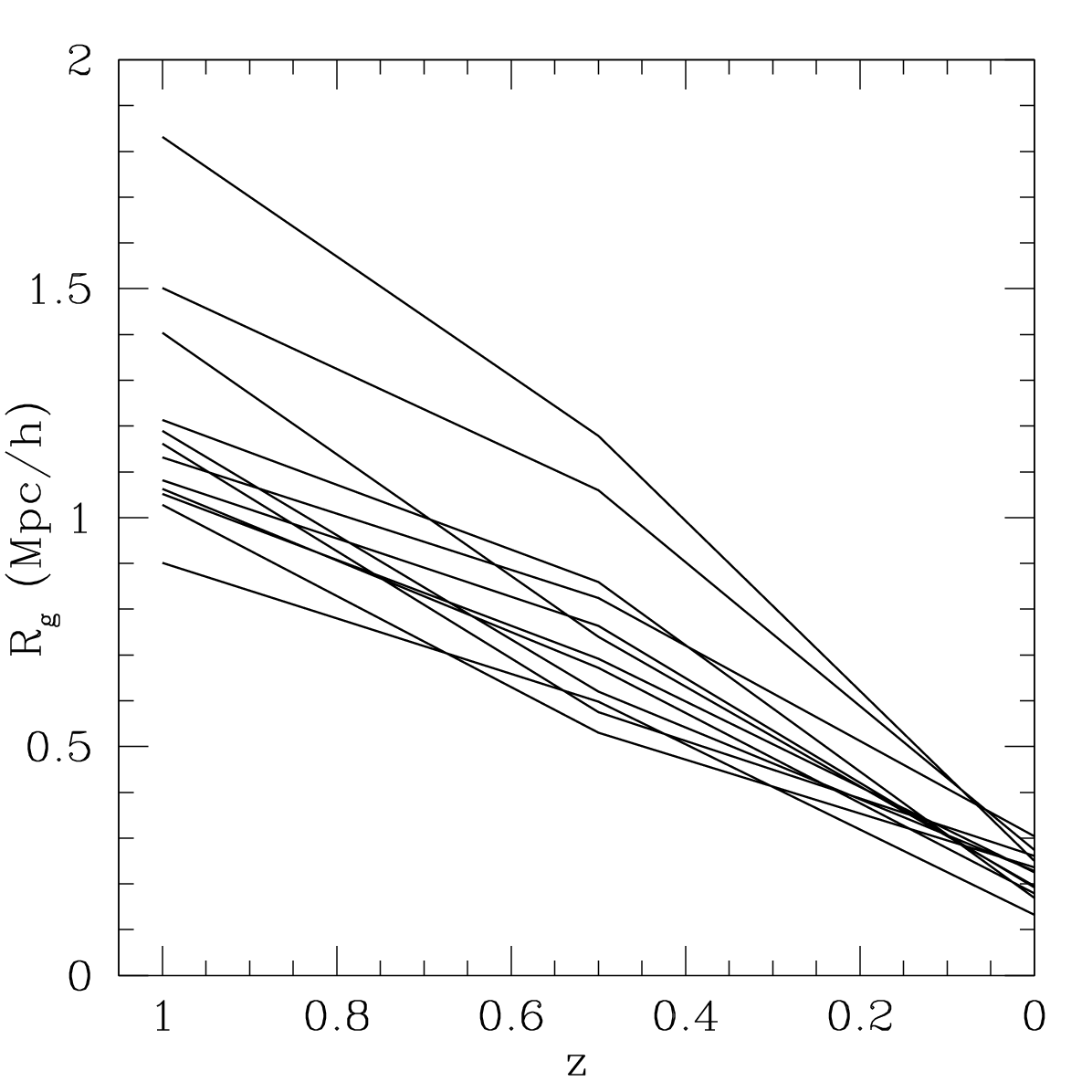}
  \vspace{-10pt}
  \caption{ Comoving group radius as a function of redshift for our CAs. The sizes were determined at $z=1.0,0.5$ and $z=0$. All CAs at $z\!=\!0$ result from a collapsing state at a higher redshift.}
  \label{fig:Revolu}
\end{figure}

In Figure~\ref{fig:groupsEvol} we show the configuration evolution in time, at different redshifts, of a particular compact group of halos in one of our simulations. As is observed,  the CA at $z=0$ results from the collapse of a loose group, and  no other normal galaxy enters a sphere of $R_{\rm max}$.  
The dynamical state of the whole group is that of a collapsing state, that has not had time to completely merge. The same trend was observed for other CAs in our simulations,  as shown graphically in Figure~\ref{fig:Revolu} were we  plot the group radius from $z=1$ to $z=0$ for our CAs.

%%%%%%%%%%%%%%%%%
%%%%%%%%%%%%%%%%%%%%%

%%%%%%%%%%%%%%%%%
%%%%%%%%%%%%%%%%%%%%%

 We also computed the crossing times of our groups of halos. The average dimensionless crossing time is $t_{\rm c}/t_{\rm H} \approx 0.03$ for CAs and $\approx 0.2$ for LAs;  where $t_{\rm c}=R_{\rm g}/\sigma$ is  a physical crossing time and $t_{\rm H}$ is the Hubble time,  and $\sigma$ is the three-dimensional velocity dispersion of the group. An observational equivalent to $t_{\rm c}$, after projecting along a line-of-sight, is typically used in observational studies (e.g. Hickson~1997).   Berlind~et~al.~(2006)  find a median value of  the dimensionless crossing time $t_{\rm cross}\approx 0.15$ for their group catalogs that includes systems of a wide range of different sizes and velocity dispersions; their crossing time relates to our deprojected value by $t_{\rm c}=4/(\pi \sqrt{3})t_{\rm cross}$. On other hand, for compact groups typically one has $t_{\rm c}/t_{\rm H}\!\sim \! 0.01$. Hence, our values for $t_{\rm c}$ are within the range found by other authors.

One might consider from our dimensionless crossing times that the halo groups are in virial equilibrium, but this is not the case. Aceves \& Vel\'azquez (2002) showed that the  value of the crossing time is not necessarily a good estimator of the dynamical state of a group.  These authors found, using dynamical studies of groups, that one can have small values  of dimensionless crossing times for collapsing groups that were clearly not in virial equilibrium; as indicated in their Fig.~4. In our cosmological simulations the average virial ratio for all associations was $\langle 2T/|U| \rangle = 0.46 \pm 0.22$ and for compact associations $0.03\pm 0.04$. From the crossing time results and  virial estimator used, it follows that the former does not appear as a solid parameter to determine if a group of the kind considered here are in virial equilibrium.  The same argument may probably apply to observational galaxy groups, but  this matter deserves further work.

\begin{figure}[t]
\centering
  \includegraphics[width=0.95\columnwidth]{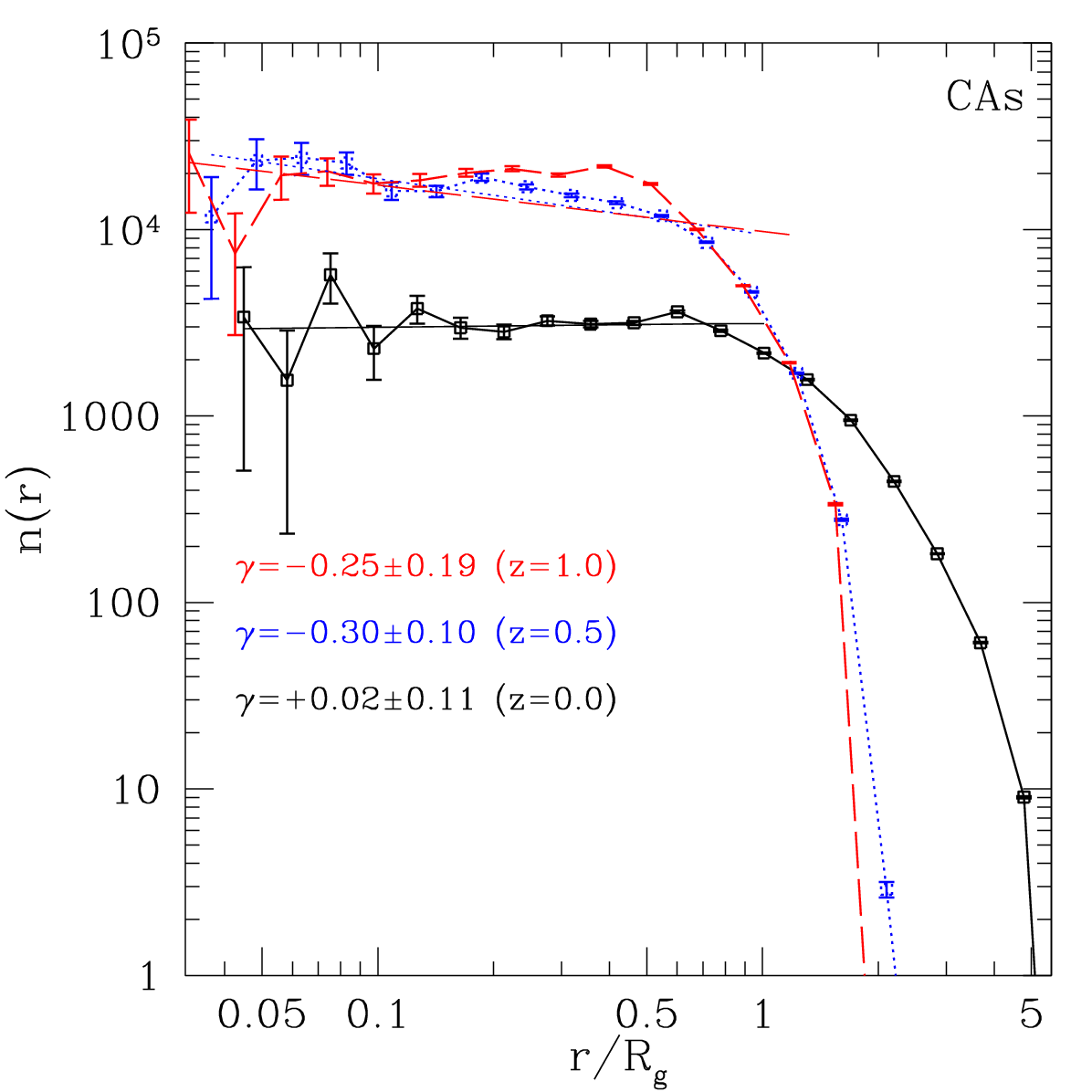}
  \vspace{-5pt}
  \caption{Evolution in time of the intragroup dark matter profile for compact associations up to $5 R_{\rm g}$ from $z=1$ to $z=0$. The inner ($r<R_{\rm g}$) logarithmic slope tends to remain constant within the uncertainties. The distribution of IG dark particles is more extended at $z=0$.}
  \label{fig:CAs-evol}
\end{figure}

In order to see if some evolutionary trends in the slopes of the IG dark matter exists, we computed the density profile for all of our CAs at three different redshifts:  $z=1.0$, $z=0.5$ and $z=0$. The results of the mass profile are shown in Figure~\ref{fig:CAs-evol}.  The average inner slope of the IG dark matter is essentially flat, within the estimated errors, from redshift $z=1$ to $z=0$. The IG dark matter appears to be more confined  within $R_{\rm g}$ at the higher redshifts than at $z=0$, as a consequence of the general collapse of the group of halos. The galaxy-size halos project the IG particles toward the external parts of the group by transferring them kinetic energy. The previous was observed by visually following the IG dark particles from $z=1$ to $z=0$ for all of our CAs. One may notice a hint of the latter by observing the IG dark particles in Figure~\ref{fig:groupsEvol}.

%%%%%%%%%%%%%%%%%
%%%%%%%%%%%%%%%%%%%%%

%%%%%%%%%%%%%%%%%%%%%%%%%%%%%%%%%%%%%%%%%%%%%%%%%%%%%%%%%%%%%%%%%%%%%%%%%
%%%%%%%%%%%%%%%%%%%%%%%%%%%%%%%%%%%%%%%%%%%%%%%%%%%%%%%%%%%%%%%%%%%%%%%%%
\section{Final Comments}\label{sec:final}
%%%%%%%%%%%%%%%%%%%%%%%%%%%%%%%%%%%%%%%%%%%%%%%%%%%%%%%%%%%%%%%%%%%%%%%%%
%%%%%%%%%%%%%%%%%%%%%%%%%%%%%%%%%%%%%%%%%%%%%%%%%%%%%%%%%%%%%%%%%%%%%%%%%

By using a set of $\Lambda$CDM cosmological simulations, with parameters in agreement with recent results from the {\sc Wmap7} observations, we have studied the distribution of dark matter in the intragroup environment of small associations of galaxy-like halos. 

In general we have found that physically well-defined halo structures, that  may resemble small groups of galaxies, have on average $\lta 40$ per cent of the total mass of the system in an intragroup medium for intermediate or loose groups, and the rest in bound halos. For compact associations  the fraction of intra-group dark matter within the group radius  ($R_{\rm g}$)  is about 20\% of the total group mass. Interestingly enough, these amounts of dark matter are comparable to the amounts ($\approx \!10-50$\%) of intracluster light found in observational studies of related astronomical systems. However, their nature appear to be different. Intracluster light comes from tidally stripped stars from galaxies, while our intragroup dark matter is the dark matter of the halos that surround galaxies. It will be of interest in the future to measure in simulations that include a baryonic component the amount of intragroup light and compare it with observations.

 Aside of the rather small amount of IG dark matter for groups, we have found that their distribution is rather flat. An {\sl average} logarithmic slope of $\gamma =0.02\pm 0.11$ in the central parts ($r\le R_{\rm g}$) of CAs is found, while for LAs $\gamma =-0.25\pm 0.15$ is obtained. In no case  a single halo was dominant or resided at the center-of-mass of our groups of dark matter halos. In some CAs we found a deficit of dark matter particles at the central parts, even after diminishing the softening radius of a cosmological simulation to $\varepsilon=1 h^{-1}\,$kpc. A better estimate of dark matter profile at such scales would certainly need to increase the number of particles in the cosmological simulations or do a re-zooming in the region of interest. However, the latter possibilities were not explored in this work. On other hand, the intragroup dark matter distribution in a particular group of halos  appears to depend on its aggregation history in a non-simple manner.

 All of our results indicate  that the distribution of dark matter in such halo associations does {\it not} follow a cuspy (e.g. a NFW) profile, contrary to what happens in individual halos formed in a $\Lambda$CDM cosmology. This results is consistent with the gravitational lensing results of Hoekstra et al.~(2001) that use groups of masses similar to the ones considered here.  Hence, what can be  called a common halo of a small galaxy group might bear in structure little resemblance to the halos of its constituent galaxies.

The results of this work may have also a relevance on works related to the dynamics of small galaxy groups. For example, our results suggest that dynamical models of the evolution of galaxies in small groups of galaxies with a large amount of intragroup matter (e.g. Athanassoula et~al.~1997) or with a cuspy profile for a common halo (e.g. Villalobos et~al.~2012) are not fully consistent  with the results presented here.
The physical compact halo associations found here are not in virial equilibrium, but in a collapsing state, so conclusions reached about the dynamical time scale for the merging of groups, based on a common homogeneous and virialized halo  (Athanassoula et~al.~1997), may not be that robust. Similarly,  the effects of the group environment modeled as a cuspy dark halo on the evolution of the discs (Villalobos et~al.~2012) may be subject to uncertainties. 
  On other hand, dynamical models of small groups where no common dark halo exists and in a collapsing state (e.g. Barnes~1985, Aceves \& Vel\'azquez~2002) would appear to be more consistent with the picture obtained here from the cosmological simulations. Researchers  of the dynamics of groups and galaxies in such environments (IAs and LAs)  may consider that about 40\% of the total mass of the system is in a common rather homogeneous dark halo, and with about 20\% when modeling compact associations.

As indicated in $\S$\ref{sec:groups} the properties of small galaxy-like groups are dependent on the algorithm chosen to determine them. Several works in the literature have constructed group catalogues in recent years (e.g. Berlind et al.~2006, Yang~et~al.~2007, Dom\'{\i}nguez-Romero~et~al.~2012). In broad terms, for example,  Berlind~et~al.~(2005) determine groups of halos (or galaxies)  in the SDSS redshift survey by looking initially for systems of galaxies that occupy a  {\sl common} dark halo; defined  as a gravitationally bound structure with a typical cosmological overdensity of 200,  that may include a galaxy, a group or cluster of galaxies. 
The groups they find   inside such overdensity are  further tested for being in virial equilibrium after obtaining small crossing times small in comparison to the Hubble time.

%%%%%%%%%%%%%%%%%
\begin{figure}[t]
\centering
  \includegraphics[width=0.9\columnwidth]{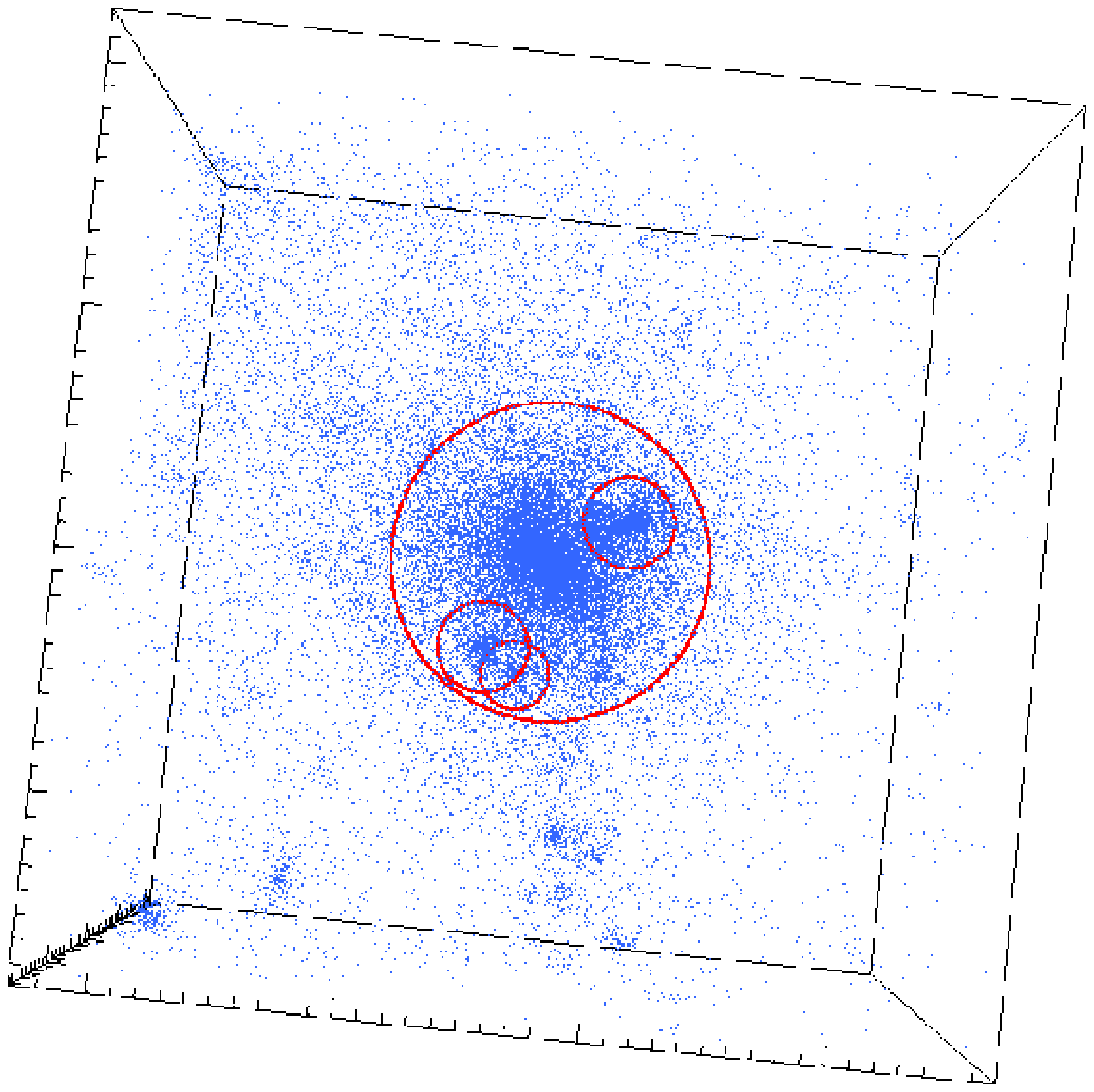}
  \caption{ Dark halo of mass $M\sim 10^{13} h^{-1}M_{\odot}$ with subhalos at $z=0$. The outer bigger red circle depicts the virial radius of the halo, and the other red circles are the virial radius of the subhalos. Box size is   $2\,h^{-1}$Mpc on each side. In this case, there is no intragroup dark matter since all dark particles belong to the main halo.}
  \label{fig:commonhalo}
\end{figure}
%%%%%%%%%%%%%%%%%%%%%

 Our groups of galaxy-size dark halos are not in virial equilibrium, even if the have very small dimensionless crossing times. We believe the difference with the above results of, for example, Berlind~et~al.~(2006) stems from the  adopted definitions of what constitutes a group. We do not use a common halo approach but agglomerations of galaxy-size dark halos to define a group. Unfortunately, we are not aware of any work on groups that study the distribution of IG dark matter so this precludes any appropriate comparison. Nonetheless, we made the following numerical exploration to have an idea of what to expect. We determined all virialized halos, defined as having an overdensity contrast of 200, with total mass $M\in [4 M_{\rm min}, 6 M_{\rm max}]\sim 10^{13} h^{-1}M_{\odot}$; similar to that of a typical group of galaxies. We found all these systems had a central halo with some subhalos; as the one shown in  Figure~\ref{fig:commonhalo}. As one would have expected, in this way of determining a group, there is a deficit of IG dark matter at the center since all dark particles belong to the main halo.

We followed here what we considered a more direct approach, probably more closely resembling standard observational methods, namely: identified halos that can host normal galaxies, look for agglomerates of them within a suitable spatial scale irrespective if they are part or not of a bigger common dark halo, and checked whether they were physically bound or not. Our physical groups of halos, although having small dimensionless crossing times, were  all in a collapsing state.  

We plan, on one hand, in the future to explore  in more detail the intragroup dark matter properties of galaxy associations obtained by different methods as the ones indicated above; however, this is out of the scope of the present work. On other hand, the dynamical fate of our groups of halos is being explored at the present.

\section*{Acknowledgments}
%%%%%%%%%%%%%%%%%%%%%%%%%%%
 
This research was funded by UNAM-PAPIIT (IN108914 \& IN104113) and CONACyT  (179662) Research Projects.  We warmly thank Mart\'{\i}n Crocce for communications regarding cosmological initial conditions, to Alexander Knebe for his help with the the AHF halo finder, and an anonymous Referee for useful comments to improve this work. We also thank Elena Jim\'enez-Bail\'on for comments and Irving \'Alvarez-Castillo, from DGCTIC-UNAM, for his help with software  matters.

%%%%%%%%%%%%%%%%%%%%%%%%%%%

%%%%%%%%%%%%%%%%%%%%%%%%%%%%%

\end{document}